\documentclass[prd,aps,twocolumn,10pt,superscriptaddress,floatfix,nofootinbib]{revtex4-2}
\usepackage{graphicx,subfigure} 
\usepackage{amsmath,amssymb,amsfonts,bm,bbm}
\usepackage{multirow}
\usepackage{epstopdf}
\usepackage{xcolor}
\usepackage{easyReview}
\usepackage{mathtools}
\usepackage{dsfont}
\usepackage{inputenc}
\usepackage{gensymb}
\usepackage[normalem]{ulem}
\usepackage[colorlinks=true,
linkcolor=magenta,
breaklinks=true,
urlcolor=blue,
citecolor=blue]{hyperref}

\graphicspath{{Figs/}}
\allowdisplaybreaks

\newcommand{\re}{\operatorname{Re}}
\newcommand{\order}[1]{\mathcal{O}\left(#1\right)}

\newcommand{\itp}{\affiliation{ Institute of Theoretical Physics, Chinese Academy of Sciences, Beijing 100190, China}}

\newcommand{\ucas}{\affiliation{School of Physical Sciences, University of Chinese Academy of Sciences, Beijing 100049, China}}

\newcommand{\uaeh}{\affiliation{\'Area Acad\'emica de Matem\'aticas y F\'isica, Universidad Aut\'onoma del Estado de Hidalgo,\\ Ciudad del conocimiento, Carr. Pachuca-Tulancingo Km. 4.5, Mineral de la Reforma, Hidalgo, 42184, Mexico}}

\newcommand{\scnt}{\affiliation{Southern Center for Nuclear-Science Theory (SCNT), Institute of Modern Physics, Chinese Academy of Sciences,\\ Huizhou 516000, China}}

\newcommand{\sxu}{\affiliation{College of Physics and Electronic Engineering, Shanxi University, Taiyuan 030006, China;}}

\begin{document}

\title{Novel method for determining the light quark mass ratio using $\eta'\to\eta \pi\pi$ decays}

\author{Adolfo Guevara} \email{adolfo\_guevara@uaeh.edu.mx}
\itp \uaeh

\author{Feng-Kun Guo} \email[Corresponding author: ]{fkguo@itp.ac.cn}
\itp \ucas \scnt

\author{Hao-Jie Jing}\email{jinghaojie@sxu.edu.cn}
\ucas \sxu 

\date{\today}

\begin{abstract}
    We propose a novel approach for extracting symmetry breaking effects from symmetry conserving three-body decays. 
    The method is based on mapping the Dalitz plot to a unit disk, and the difference of the disk distributions of two related decays isolates purely symmetry breaking effects. We demonstrate this method by extracting the fundamental parameter $Q$, an isospin breaking ratio of light quark masses defined as $Q^2\equiv (m_s^2-\hat m^2)/(m_d^2-m_u^2)$ with $\hat m$ the average of up and down quark masses, from the 
    decays $\eta'\to\eta\pi^+\pi^-$ and $\eta'\to\eta\pi^0\pi^0$. 
    With the Dalitz plot distributions for these two decays reported by BESIII, we illustrate the method and obtain $Q=22.5\pm1.0$, which is consistent with previous determinations and has a comparable uncertainty. With the full BESIII data set, which is eight times larger than the one used here, a more precise determination of $Q$ should become possible. This promising and novel method can be generalized to other three-body decays to extract symmetry breaking effects.
\end{abstract}

\maketitle
	
\section{Introduction}

As experimental particle physics advances towards increasingly precise measurements, further phenomenological developments are essential to interpret the results accurately. It is in this context that the Dalitz plot method~\cite{Dalitz:1953cp,Fabri:1954zz} was developed, and it has since become a widely used tool to analyze three-body decays in modern particle physics. While this method has proven valuable for visualizing various phenomena, its utility can be further enhanced through targeted improvements. One notable feature of the Dalitz plot is that its area and shape depend on the masses of the particles involved.
Here, we propose a novel approach: by applying a suitable change of variables, the Dalitz plot can be mapped onto a unit disk. This unit disk can then be discretized into bins. For two three-body decays related by a symmetry, their normalized Dalitz plot distributions can both be mapped onto unit disks. By taking the bin-by-bin difference between these distributions, a new unit disk distribution is obtained, which isolates symmetry-breaking effects. Then symmetry breaking parameters can be extracted from such a distribution.

As an example, we apply this method to the $\eta'\to\eta\pi\pi$ decays to extract one important parameter of the Standard Model, the double ratio $Q^2\equiv(m_s^2-\hat m^2)/(m_d^2-m_u^2)$~\cite{Leutwyler:1996sa} (corresponding to $1/\kappa$ in Ref.~\cite{Gasser:1984pr}) of light quark masses, where $\hat m\equiv (m_u+m_d)/2$ is the averaged up and down quark masses.
In this way, one can make use of the whole Dalitz plot information to extract the symmetry breaking effects, well beyond using only the branching fractions, $\mathcal{B}(\eta'\to\eta\pi^+\pi^-)=(42.5 \pm 0.5) \%$ and $\mathcal{B}(\eta'\to\eta\pi^0\pi^0)=(22.4 \pm 0.5) \%$~\cite{ParticleDataGroup:2024cfk}.
Previous attempts to access light-quark mass ratios from $\eta'$ decays focused on rate ratios involving the isospin-forbidden decay $\eta'\to\pi^0\pi^+\pi^-$~\cite{Borasoy:2006uv}. Our approach instead uses two symmetry-conserving channels and exploits the full differential information in their Dalitz plot distributions.
As will be demonstrated, applying this method to the published BESIII data on the decays $\eta'\to\eta\pi^+\pi^-$ and $\eta'\to\eta\pi^0\pi^0$ reported in Ref.~\cite{BESIII:2017djm} allows for a determination of $Q^2$ with a 5\% uncertainty. A significantly more precise determination will be achievable once the full BESIII $\eta'$ data set~\cite{BESIII:2022tas}, which is eight times larger than that used in Ref.~\cite{BESIII:2017djm}, is released. Thus, the unit disk mapping proposed in this paper represents a promising and novel approach for precisely extracting the quark mass ratio. For a comprehensive review of precision tests of fundamental physics using $\eta$ and $\eta'$ decays, we refer to Ref.~\cite{Gan:2020aco}.

\section{Isospin breaking in $\eta'\to\eta\pi\pi$ decays}

    Isospin symmetry requires that the $u$ and $d$ quarks are identical. Nevertheless, their electric charges and masses differ, leading to two sources of isospin-breaking effects:  electromagnetic interactions and the mass difference between the up and down quarks. 
    Both contributions can be systematically studied using chiral perturbation theory (ChPT)~\cite{Weinberg:1978kz,Gasser:1983yg,Gasser:1984gg} with virtual photons~\cite{Urech:1994hd}. It has been shown that electromagnetic corrections to the extraction of the $Q$ parameter from isospin breaking (IB) $\eta\to3\pi$ decays are negligibly small, at the percent level of the isospin-breaking effects stemming from the quark mass difference~\cite{Ditsche:2008cq}.
    For the processes considered in here, nevertheless we will consider electromagnetic corrections in this work.
    The isospin-breaking-induced threshold cusp at the $\pi^+\pi^-$ threshold in the $\pi^0\pi^0$ invariant mass distribution of the $\eta'\to\eta\pi^0\pi^0$ has been studied in Ref.~\cite{Kubis:2009sb} using a nonrelativistic effective field theory framework.

    Since the reactions under study involve the $\eta'$ meson, a suitable framework to describe the decay amplitudes is ChPT with large $N_c$~\cite{Leutwyler:1996sa,Kaiser:2000gs}, with $N_c$ the number of quark colors. This framework employs a triple expansion in powers of light quark masses, powers of momentum, and powers of $1/N_c$ with the power counting $\partial_\mu=\mathcal{O}(\sqrt{\delta})$, $m_q=\mathcal{O}(\delta)$, $1 / N_c=\mathcal{O}(\delta)$. 
    However, it has been found that the chiral perturbative amplitude alone up to the next-to-leading order (NLO) does not yield the correct values for the Dalitz plot parameters~\cite{Fariborz:1999gr}.
    Given that the $\pi\pi$ invariant mass for the $\eta'\to \eta\pi\pi$ decays can reach up to 0.41~GeV, it is essential to account for the $S$-wave $\pi\pi$ final state interaction in a nonperturbative manner to incorporate the $f_0(500)$ contributions, as demonstrated for $\eta\to3\pi$ decays~\cite{Anisovich:1996tx}.
    This has been done in Refs.~\cite{Borasoy:2005du,Borasoy:2006uv,Escribano:2010wt,Gonzalez-Solis:2018xnw} using unitarized ChPT and in Refs.~\cite{Isken:2017dkw,Akdag:2021efj} using dispersion relations.
    Here we follow the description outlined in Ref.~\cite{Escribano:2010wt}, where an NLO analysis within large $N_c$ ChPT was performed including $\pi\pi$ rescattering effects via the $N/D$ unitarization method~\cite{Chew:1960iv,Oller:1998zr}.

\begin{figure}[tb]
        \centering
        \includegraphics[width=\linewidth]{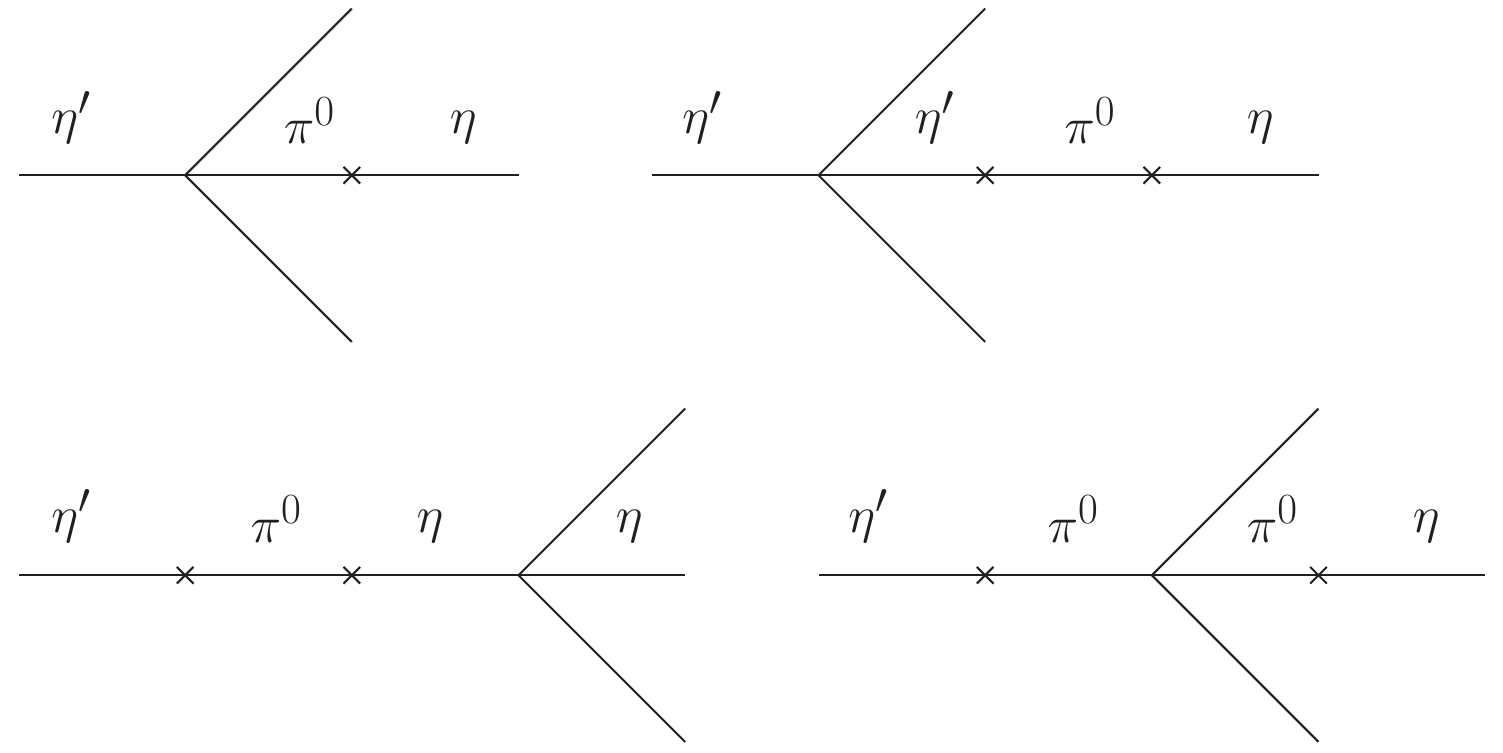}
        \caption{Feynman diagrams of the contributions with two $\Delta I=1$ insertions. Note that the $\eta'\to3\pi$ vertex in the first diagram is also isospin breaking. The lines that are not tagged correspond to either charged or neutral pions, depending on the decay channel.}
        \label{fig:IB_FeynmanD}
    \end{figure}

    Because the $\eta'\to\eta\pi^+\pi^-$ and $\eta'\to\eta\pi^0\pi^0$ decay processes are allowed by isospin symmetry (the isospin conserving amplitudes as given in Ref.~\cite{Escribano:2010wt} are provided in Appendix~\ref{app:ICamp}), the IB effects in the decay amplitudes must involve at least two $\Delta I=1$ vertices proportional to $m_d-m_u$ (shown in Fig.~\ref{fig:IB_FeynmanD}) or one $\Delta I=2$ vertex, which can be generated by electromagnetic operators at $\mathcal{O}(e^2p^2)$ in ChPT with virtual photons~\cite{Urech:1994hd, Knecht:1997jw}. 

    The LO ($\order{\delta^0}$) Lagrangian and relevant terms in the NLO ($\order{\delta}$) Lagrangian of the large $N_c$ ChPT contributing to IB effects in the reactions under study read~\cite{Kaiser:2000gs,Escribano:2010wt,Urech:1994hd} 
    \begin{align}\label{eq:L_2}
        \mathcal{L}_{(0)}=&\,
        \frac{F^2}{4}\left(\langle\partial_\mu U^\dagger\partial^\mu U\rangle+
        \langle U^\dagger\chi+\chi^\dagger U\rangle \right) -\frac{1}{2} M_0^2 \eta_1^2, \notag\\
        \mathcal{L}^{\rm IB}_\text{(1)}=&\, L_5\langle\partial_\mu U^\dagger\partial^\mu U(U^\dagger\chi +\chi^\dagger U)\rangle \notag\\ 
        &+ L_8\langle U^\dagger\chi U^\dagger\chi+\chi^\dagger U\chi^\dagger U\rangle -i\frac{F\Lambda_2}{2\sqrt{6}}\eta_1\langle U^\dagger\chi-\chi^\dagger U\rangle ,\nonumber\\
        \mathcal{L}_\text{em}=&\,K_6\left\langle D^\mu U^\dagger D_\mu U\mathcal{Q}U^\dagger\mathcal{Q}U+D^\mu UD_\mu U^\dagger\mathcal{Q}U\mathcal{Q}U^\dagger\right\rangle+\notag\\
        &+K_9\left\langle(\chi^\dagger U+U^\dagger \chi+\chi U^\dagger + U\chi^\dagger)\mathcal{Q}^2\right\rangle\notag\\
        &+K_{10/11}\left\langle(\chi^\dagger U\pm U^\dagger \chi)\mathcal{Q}U^\dagger\mathcal{Q}U \right. \notag\\
        &\left. +(\chi U^{\dagger}\pm U\chi^\dagger)\mathcal{Q}U\mathcal{Q}U^\dagger\right\rangle,
    \end{align}
    where $\cal L_\text{em}$ contains the relevant terms with electromagnetic contributions at $\order{e^2p^2}$. $U$ contains the pseudo-Nambu-Goldstone bosons, $F$ is the pion decay constant in the chiral limit, the singlet $\eta_1 = \left(C_0\eta+C_0'\eta'\right)$ is a mixture of $\eta$ and $\eta'$, $M_0$ is the U(1)$_A$ anomaly contribution to the $\eta_1$ mass, $L_5$, $L_8$ and $\Lambda_2$ are low-energy constants (LECs), $\chi=2B_0 \mathcal{M}$, $B_0$ is related to the quark condensate, $\mathcal{M}$ is the light quark mass matrix, $\mathcal{M}= \text{diag}(m_u,m_d,m_s)=\mathds{1}(m_u+m_d+m_s)/3+ \lambda_3(m_u-m_d)/{2}+\lambda_8(\hat m-m_s)/{\sqrt{3}}$,
    with $\lambda_{3,8}$ the Gell-Mann matrices, and $\mathcal{Q}$ the quark charge matrix.
    As shown in Ref.~\cite{Osborn:1970nn}, the quark-mass induced IB effects come from operators with $I=1$, which is the $\lambda_3$ term of the quark-mass matrix. 
    This implies that the $m_u-m_d$ insertions considered here will change isospin by one unit. Electromagnetic operators can also generate $\Delta I=2$ contributions, as discussed above.

    The IB contributions to the $\eta'$ decays including both up-down quark mass difference and electromagnetic corrections read 
    \begin{align}\label{eq:all_IB_amplitudes}
    \mathcal{M}^{\rm IB}_{\pi^0}=&
    \left[B_0(m_u-m_d)\right]^2\left\{\left[\left(\frac{\kappa C_\eta C'_\eta}{\Delta_{\eta'\pi}}-\frac{C_{\eta'}C'_{\eta'}}{\kappa\Delta_{\eta\pi}}\right)\frac{1}{\Delta_{\eta'\eta}}\right.\right.\nonumber\\
    &\left.-2\frac{C_\eta}{\Delta_{\eta\pi}}\left(\frac{C_{\pi}}{\Delta_{\eta\pi}}+\frac{C'_{\pi}}{\kappa\Delta_{\eta'\pi}}\right)+3\frac{C'_{\eta'}C_\eta}{C_qC_q'\Delta_{\eta'\pi}\Delta_{\eta\pi}}\right]\nonumber\\
    &\left.\frac{}{}\mathcal{M}_{\eta'\to\eta\pi\pi}^{\rm ChPT}+3\frac{C_0C_\eta}{F^2\Delta_{\eta\pi}} + \frac{64L_8 C_q C_q'}{F^4} \right\}+\mathcal{M}^\text{em}_{\pi^0}
    ,\\
    \mathcal{M}^{\rm IB}_{\pi^\pm}=&
    \,\left[B_0(m_u-m_d)\right]^2\left[\left(\frac{\kappa C_\eta C'_\eta}{\Delta_{\eta'\pi}}-\frac{C_{\eta'}C'_{\eta'}}{\kappa\Delta_{\eta\pi}}\right)\frac{\mathcal{M}_{\eta'\to\eta\pi\pi}^{\rm IC}}{\Delta_{\eta'\eta}}\right.\nonumber\\
    &\left.+\frac{C'_{\eta'}C_\eta \mathcal{M}_{\pi^0\to\pi^+\pi^-\pi^0}^{\rm IC}}{\Delta_{\eta'\pi}\Delta_{\eta\pi}}+\frac{C_\pm C_\eta}{3F^2\Delta_{\eta\pi}}\right]+\mathcal{M}^\text{em}_{\pi^\pm}, \notag 
    \end{align}
    where the subscript $\pi^{0}$ ($\pi^\pm$) in the amplitude refers to the decay channel with two neutral (charged) pions, $\mathcal{M}_{\eta'\to\eta\pi\pi}^{\rm IC}$ and $\mathcal{M}_{\pi^0\to\pi^+\pi^-\pi^0}^{\rm IC}$ are the isospin conserving amplitudes, $\Delta_{\varphi_1\varphi_2}\equiv M_{\varphi_1}^2 - M_{\varphi_2}^2$, and 
    \begin{align}
        C^{(\prime)}_{\phi} &= -C_q^{(\prime)}-\frac{\sqrt{2}\Lambda_2}{3}C_1^{(\prime)}+\frac{4L_5}{F^2}C_q^{(\prime)}M^2_\phi-16\frac{L_8C_q^{(\prime)}}{F^2},\nonumber\\ 
        C_\pm&=C_q'+\frac{\sqrt{2}\Lambda_2}{3}C_1'+12\frac{L_5}{F^2}C_q'\Delta_{\eta\pi}+32\frac{L_8C_q'}{F^2}, \nonumber\\
        C_0 &=C_q'+\frac{\sqrt{2}\Lambda_2}{3}C_1'+32\frac{L_8C_q'}{F^2},\nonumber \\
        C_1^{(\prime)}&=\sqrt{2}C_q^{(\prime)}\mp C_s^{(\prime)},\ \quad \kappa=\frac{C_q}{C'_q} ,\nonumber\\
        \mathcal{M}^\text{em}_{\pi^0}\,  &\,=\frac{20e^2C_qC_q'M_\pi^2}{9F^2}(K_9+K_{10}),\nonumber\\
        \mathcal{M}^\text{em}_{\pi^\pm}\, &\, =\frac{4e^2C_qC_q'}{9F^2}\left[M_\pi^2K_m+K_6(s-M_{\eta'}^2-M_\eta^2)\right],\nonumber\\
        K_m\, &\, =5K_9+23K_{10}+18K_{11}
    \end{align}
    with $C_q^{(\prime)}$ and $C_s^{(\prime)}$ related to the $\eta$-$\eta'$ mixing angles in the two mixing-angle scheme~\cite{Schechter:1992iz,Kisselev:1993,Kaiser:2000gs,Feldmann:1998vh,Feldmann:1998sh} (Ref.~\cite{Guevara:2018rhj} for explicit expressions) and $s=(p_{\pi^{+}}+p_{\pi^{-}})^2$. 

    The $Q$ parameter will be extracted from the IB unit disk distribution using the following relation~\cite{Gasser:1984pr, Gan:2020aco},
    \begin{equation}
        B_0(m_d-m_u)=\frac{M_K^2}{M_\pi^2}\frac{M_K^2-M_\pi^2}{Q^2} + \order{m_q^3}.
    \end{equation} 

  We account for $\pi\pi$ final state interactions with the $N/D$ unitarization method~\cite{Chew:1960iv,Oller:1998zr} as of Ref.~\cite{Escribano:2010wt}, and the $\eta\pi$ rescattering ($t$- and $u$-channels) in these decays is negligible~\cite{Schneider:2009rz,Kubis:2009vu} (thus the contribution from the $a_0(980)$ resonances that couple to $\eta\pi$ and located well beyond the phase space limit is also expected to be negligible).  Details of these partial-wave amplitudes can be found in Ref.~\cite{Escribano:2010wt}.

\section{Unit disk mapping}

Consider a three-body decay with an initial particle of mass $m$ and final-state particles of masses $m_1$, $m_2$ and $m_3$. Let $m_{12}$ and $m_{23}$ denote the invariant masses of particles 1 and 2, and particles 2 and 3, respectively.
The Dalitz plot boundary is determined by solving the equation $B\left(m_{12}^2, m_{23}^2\right)=\pm 1$ with $B\left(m_{12}^2, m_{23}^2\right)$ given by~\cite{ParticleDataGroup:2024cfk} 
\begin{equation}
  B\left(m_{12}^2, m_{23}^2\right)\equiv \frac{2 E_2^* E_3^*+m_2^2+m_3^2-m_{23}^2}{2 q_2^* q_3^*},
  \label{eq:B_function}
\end{equation}
where $E_{2(3)}^*$ and $q_{2(3)}^*$ represent the energy and the magnitude of the three-momentum of particle $2(3)$ in the center-of-mass (c.m.) frame of particles 1 and 2.

\begin{figure}
    \centering
    \includegraphics[width=0.9\linewidth]{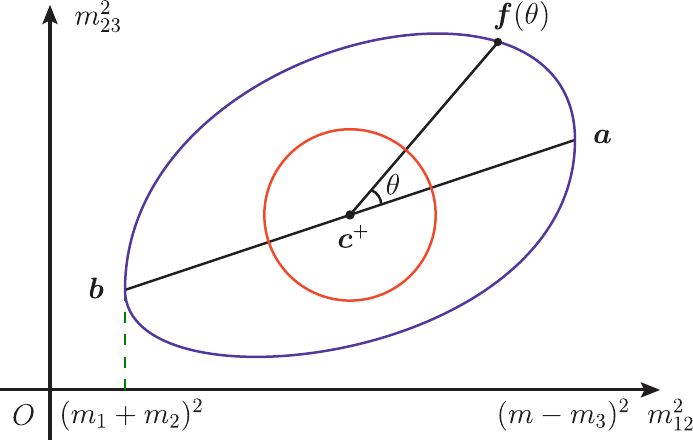}
    \caption{Illustration of the mapping from a Dalitz plot $B$ to a unit disk $D$, where $m$ is the mass of initial state particle, and $m_i~(i=1,2,3)$ are masses of final state particles. }
    \label{fig:mapping}
\end{figure}

A one-to-one mapping from a unit circle $D$ to the boundary of the Dalitz plot $B$, denoted as $\bm{f}: D \rightarrow B$, can be constructed as illustrated in Fig.~\ref{fig:mapping}. The mapping is defined by
\begin{equation}
\bm{f}: \theta \mapsto\left(m_{12}^2, m_{23}^2\right) = \left(L(\theta), R(\theta)\right),
\end{equation}
where $L(\theta)$ is the solution to the equation
\begin{equation}\label{eq:Boundary_Soln}
  B\big(m_{12}^2, S\left(m_{12}^2,\theta\right)\big)=\left\{\begin{array}{cl}1 & \text { for } \theta \in[0, \pi) \\ -1 & \text { for } \theta \in[\pi, 2 \pi)\end{array}\right.
\end{equation}
with respect to $m_{12}^2$, and $R(\theta)=S\big(L(\theta),\theta\big)$. Here, $S(m_{12}^2,\theta)$ represents the value of $m_{23}^2$ for a given $m_{12}^2$ and $\theta$.
The center of the Dalitz plot $\bm{c}^+ \equiv (\bm{a}+\bm{b})/2$ is mapped to the origin of the unit circle,
where the two-component vectors $\bm{a}$ and $\bm{b}$ are the two endpoints of the Dalitz plot along the $m_{12}^2$ axis. 
The one-to-one mapping from the entire Dalitz plot to the unit disk, with $D$ as its boundary, can then be constructed (technical details of this mapping are provided in Appendix~\ref{app:mapping}).

In constructing the conventional Dalitz plot, the limits in the integral for the total width depend on the masses of the particles involved in the process. When changing the integration variables to those of the unit disk, the dependence on the masses will be transferred to the Jacobian 
\begin{equation}
    |J|=r\left(\bm{f}(\theta)-\bm{c}^+\right)^2 
\end{equation}
of the transformation, where $r$ is the radial coordinate of the unit disk. The three-body partial width can then be expressed as
\begin{align}\label{eq:Jacobian}
        \Gamma&=\int dm_{12}^2dm_{23}^2\frac{d^2\Gamma}{dm_{12}^2dm_{23}^2}\nonumber\\
        &={\int_{0}^1 dr\int_{0}^{2\pi}d\theta\left|J\right|\frac{d^2\Gamma}{dm_{12}^2dm_{23}^2}}.
\end{align}
Thus, the difference in partial widths for the $\eta'\to\eta\pi^0\pi^0$ and $\eta'\to\eta\pi^+\pi^-$ decays is 
\begin{equation}
        2\Gamma_{\pi^0}-\Gamma_{\pi^\pm}=\int_{0}^1 dr\int_{0}^{2\pi}d\theta\, d\Gamma'_\text{diff},
\end{equation}
where the difference width is defined as
\begin{equation}
        d\Gamma_\text{diff}'(r,\theta)\equiv |J_{\pi^0}|\frac{ d^2\Gamma_{\pi^0}}{dm_{12}^2dm_{23}^2}(r,\theta) - |J_{\pi^\pm}| \frac{ d^2\Gamma_{\pi^\pm}}{dm_{12}^2dm_{23}^2}(r,\theta).
\end{equation}
The factor $\Delta_J(\theta)\equiv 1-{\left|J_{\pi^\pm}\right|}/{\left|J_{\pi^0}\right|}$ signifies the difference between the unit disk mapping of the Dalitz plot regions for the decays. 
It is also an IB effect and is proportional to $M_{\pi^\pm}^2-M_{\pi^0}^2$ to a very good approximation. Specifically, $\Delta_J(\theta)$ is independent of $r$ and numerically in the range of $[10.2,13.3]\%$ (for $\theta=0$, one has $\Delta_J(0)\simeq {8\left(M_{\pi^\pm}^2-M_{\pi^0}^2\right)}/{\left[\left(M_{\eta^{\prime}}-M_\eta\right)^2-4 M_{\pi^0}^2\right]}$).

The above expression involves a large background term $(|J_{\pi^0}|-|J_{\pi^\pm}|)|\mathcal{M}_\textrm{IC}|^2$, where $\mathcal{M}_\textrm{IC}$ is the isospin conserving (IC) amplitude. 
 Therefore, instead of using the previous expression, to generate the difference disk we use 
 \begin{equation}
        d\Gamma_\text{diff}(r,\theta)\equiv \frac{ d^2}{dm_{12}^2dm_{23}^2}\left[ \Gamma_{\pi^0}(r,\theta)-\Gamma_{\pi^\pm}(r,\theta)\right].
\end{equation}
Since the theoretical decay width $d^2\Gamma/drd\theta$ must coincide with that obtained from experimental data, each Jacobian involved in changing to $(m_{12}^2,m_{23}^2)$ will be included as a factor of each differential decay width obtained from experimental data.

The decay amplitude for each decay contains both the IC and IB contributions. Since the former is the same for both decays,
 we have 
\begin{align}
d\Gamma_\text{diff}(r,\theta) = 2\re\!\left({\mathcal{M}_{\rm IC}^*}\mathcal{M}_{\rm IB}\right) + \order{Q^{-8}},
        \label{eq:dGamma_diff}
\end{align}   
where $\mathcal{M}_\text{IC}$ is the IC amplitude, and $\mathcal{M}_\text{IB}$ is the difference between the $\eta'\to\eta\pi^0\pi^0$ and $\eta'\to\eta\pi^+\pi^-$ amplitudes. 
Notice that all the amplitudes have been unitarized to account for the $\pi\pi$ rescattering as in Ref.~\cite{Escribano:2010wt}.

The term in Eq.~\eqref{eq:dGamma_diff} is proportional to $1/Q^4$, which allows it to probe the $Q$ parameter at the same level of sensitivity as the $\eta\to3\pi$ decays.

\section{Extraction of $Q$}
 
\begin{figure}[tb]
    \centering
    \includegraphics[scale=0.375]{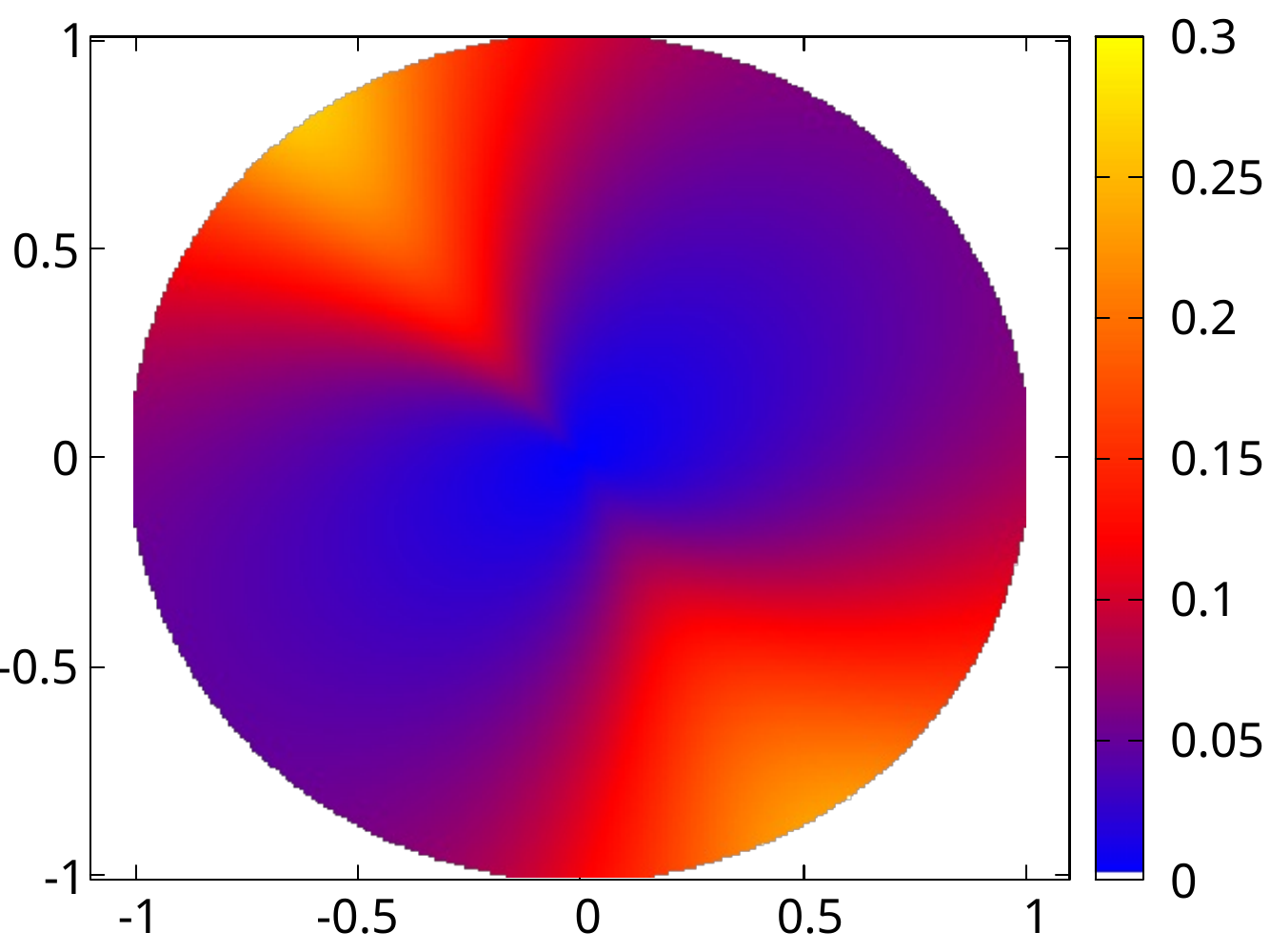}
    \caption{Unnormalized unit disk of the $\eta'\to\eta\pi^+\pi^-$ decay generated with the BESIII model~\cite{BESIII:2017djm}. 
    }
 \label{fig:Neut_disk}
\end{figure} 

    We utilize the Dalitz plot distributions measured by BESIII, which are based on samples of $3.51\times10^5$ and $5.62\times10^4$ events for  $\eta'\to\eta\pi^+\pi^-$ and $\eta'\to\eta\pi^0\pi^0$, respectively, obtained from $1.31\times10^9$ $J/\psi$ events~\cite{BESIII:2017djm}. These samples amount to only about 1/8 of the full BESIII data set~\cite{BESIII:2022tas}. 
    The Dalitz plot distribution is parameterized by an expansion around the center of the Dalitz plot as~\cite{BESIII:2017djm}  
    \begin{equation}
        \frac{d^2\Gamma}{dXdY}=N(1+aY+bY^2+dX^2+...),
        \label{eq:dalitz}
    \end{equation}
    with the Dalitz plot distribution parameters $a,b$ and $d$, and expansion variables $X$ and $Y$
    \begin{align}
        X_{\pi^\pm}&\equiv\frac{\sqrt{3}(T_{\pi^+}-T_{\pi^-})}{M_{\eta'}-M_\eta-2M_{\pi^\pm}},\quad 
        X_{\pi^0}\equiv\frac{\sqrt{3}|T_{\pi^0_1}-T_{\pi^0_2}|}{M_{\eta'}-M_\eta-2M_{\pi^0}}, \notag \\
        Y&\equiv\frac{M_\eta+2M_{\pi f}}{M_{\pi f}}\frac{T_\eta}{M_{\eta'}-M_\eta-2M_{\pi f}}-1,\label{eq:Dalitz_Y}
    \end{align}
    where $T_i$ is the kinetic energy of particle $i$, and $M_{\pi f}$ corresponds either to the final state charged or neutral pion mass depending on the decay channel. The normalization factor $N$ in the BESIII model~\cite{BESIII:2017djm} can be obtained from the branching fraction for each decay channel. The unit disk distribution  generated using the BESIII model for the Dalitz plot distribution of $\eta'\to\eta\pi^+\pi^-$ is shown in Fig.~\ref{fig:Neut_disk}.
  
    \begin{figure}[tb]
        \centering
        \includegraphics[scale=0.375]{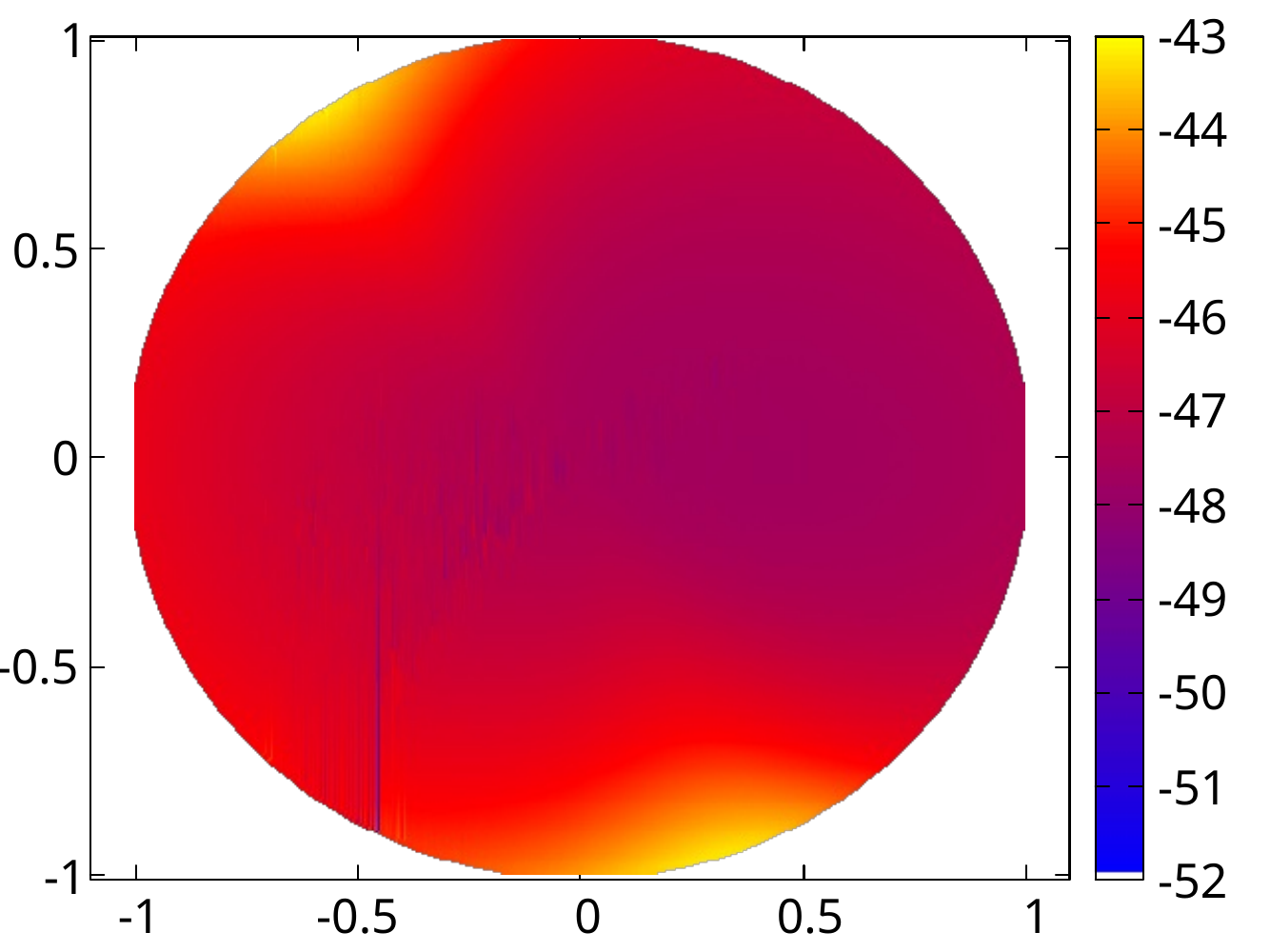}
        \caption{Unit disk of the difference between the charged and neutral decays generated using the BESIII data in Ref.~\cite{BESIII:2017djm}.}
        \label{fig:Diff_disk}
    \end{figure}
   
    We divide the unit disk into bins in the following way: we partition the circumscribed square of the unit circle into $10^4$ identical bins and then consider those that lay inside the unit disk to be the ones whose coordinates accomplish the relation $\sqrt{x^2+y^2}\leq1$. In this way, we obtain a total of 7837 bins (a large number of bins, in particular along the edge of the unit disk, is necessary to reduce the error from the mismatch between the area of the unit disk and that of the binned square; in this case, it is 0.2\% away from the exact relation between the areas of the unit circle and its circumscribed square).
    The unit disk distribution for each decay is obtained from the constructed mapping using a Monte Carlo run with 7000 points per bin, which generates randomly the $a,b$ and $d$ parameters with a normal distribution according to their mean values, errors and correlations from the BESIII measurements~\cite{BESIII:2017djm} (listed in Appendix~\ref{app:parameters}). 
    The binned disk representation is a deterministic transformation of these experimental inputs; binning does not increase the number of independent degrees of freedom, but serves to propagate the correlated BESIII uncertainties to the disk distributions (standard Monte-Carlo replica propagation of a covariance matrix, see e.g. Refs.~\cite{NNPDF:2014otw,DelDebbio:2021whr}).
    Furthermore, we also fit the overall normalization of each disk to account for the difference in theoretical and experimental partial widths.
    Then, subtracting the differential decay width of one disk from the other, bin by bin, we obtain a unit disk distribution of the difference, as shown in Fig.~\ref{fig:Diff_disk}. 

    Since the $3L_2+L_3$ term compellingly dominate over the other NLO terms ($L_5$, $L_8$ and $\Lambda_2$) in the IC amplitudes~\cite{Escribano:2010wt}, we keep $L_2$,
    $L_3$ and $Q$ as free parameters,\footnote{Although the combination $3L_2+L_3$ for the IC amplitude appears instead of $L_2$ and $L_3$ alone, it is not the case for the $\mathcal{M}_{\pi\pi\to\pi\pi}$ amplitude used for rescattering effects.} along with the normalization of each disk, while setting $L_5= 2.1\times 10^{-3}$, $L_8= 0.8\times10^{-3}$, and $\Lambda_2 = 0.3$ from {\it{}fit 3} of Ref.~\cite{Escribano:2010wt}. As for the $K_i$ values, we take $K_6=2.8\times10^{-3}$, $K_9=-1.3\times10^{-3}$~\cite{Bijnens:1996kk}, $K_{10}=4.0\times10^{-3}$ and $K_{11}=1.3\times10^{-3}$ evaluated at a scale $\mu=770$~MeV following Ref.~\cite{Ditsche:2008cq}.
    We obtain 
    \begin{align}
        L_2&= 0.97(5)\times 10^{-3},
        \quad L_3 = -4.3(2)\times10^{-3}, \nonumber\\
        Q &= 22.5(1.0),\quad N_{\pi^\pm}=130(3), \quad N_{\pi^0}=70(5), \label{eq:values}
    \end{align}
    with $\chi^2/\text{dof} = 18792/7832$. Here the denominator refers to the number of bins minus the number of fitted parameters and provides an effective goodness-of-fit indicator for the binned representation; it should not be interpreted as the number of independent experimental inputs, which are the Dalitz-plot parameters with their covariance matrices. 
    All uncertainties have been scaled by the Birge factor~\cite{Birge:1932hlj,ParticleDataGroup:2024cfk} $\sqrt{\chi^2/{\rm dof}}=1.55$ 
    to account for the goodness of the fit.\footnote{If the fit is done again but neglecting the electromagnetic corrections, one obtains $Q=22.1(1.1)$, $L_2=1.01(3)\times10^{-3}$ and $L_3=-4.4(1)\times10^{-3}$ with a $\chi^2/\text{dof}=18652/7832$. The results are compatible with those in Eq.~\eqref{eq:values}, showing that the electromagnetic corrections are marginal though larger than in $\eta\to3\pi$~\cite{Ditsche:2008cq}.}
    We have checked that if we use other values for $L_5$, $L_8$  and $\Lambda_2 $ from the other fits in the same reference instead, the result on $Q$ remains unchanged. As shown in Table~\ref{tab:Q_comparison}, the $Q$ value determined in this way is compatible with previous phenomenological~\cite{Anisovich:1996tx,Kambor:1995yc,Bijnens:2007pr,Kampf:2011wr,Colangelo:2010et,Colangelo:2016jmc,Colangelo:2018jxw,Albaladejo:2017hhj} (see Ref.~\cite{Gan:2020aco} for a review) and lattice~\cite{FlavourLatticeAveragingGroupFLAG:2024oxs} determinations. 
    The uncertainty on $Q$ without the Birge factor ($\pm0.7$) is also comparable to the ones obtained from $\eta\to 3\pi$ in previous studies~\cite{Anisovich:1996tx,Kampf:2011wr,Colangelo:2018jxw,Albaladejo:2017hhj}.
    Therefore, once the full BESIII data set (8 times larger than the one used here) is analyzed, the uncertainty on $Q$ will be significantly reduced.

    \begin{table}[tb]
        \centering
        \caption{Comparison between our determination of the $Q$ parameter and previous determinations. Note that the uncertainty of our result has been rescaled by multiplying by the Birge factor $\sqrt{\chi^2/{\rm dof}}=1.55$, which takes into account the goodness of the fit.}
        \begin{ruledtabular}
        \begin{tabular}{l l}
           $Q$ &  Refs. \\\hline
           24.3&  from Dashen's theorem \cite{Dashen:1969eg}\\
           $22.7\pm0.8$ &  A. V. Anisovich \& H. Leutwyler \cite{Anisovich:1996tx}\\
           $23.1\pm0.7$ &  K. Kampf {\it et al.} \cite{Kampf:2011wr}\\
           $22.1\pm0.7$ &  G. Colangelo {\it et al.} \cite{Colangelo:2018jxw}\\
           $21.50\pm0.97$ &  M. Albaladejo \& B. Moussallam \cite{Albaladejo:2017hhj}\\
           $22.4\pm0.3$ & D. Stamen {\it et al.} \cite{Stamen:2022uqh}\\
           $23.3\pm0.5$ &  FLAG $(N_f=2+1)$ \cite{FlavourLatticeAveragingGroupFLAG:2024oxs}\\
           $22.5\pm0.5$ &  FLAG $(N_f=2+1+1)$ \cite{FlavourLatticeAveragingGroupFLAG:2024oxs}\\
           $22.5\pm1.0$ &  this work
        \end{tabular}
    \end{ruledtabular} 
        \label{tab:Q_comparison}
    \end{table}

    \section{Conclusions}

    In this paper, a novel method for extracting symmetry breaking information from symmetry related three-body reactions is proposed.  This method extracts symmetry breaking effects from symmetry conserving three-body decays by constructing unit disk distributions. 
    We apply the method to the decays $\eta'\to\eta\pi^0\pi^0$ and $\eta'\to\eta\pi^+\pi^-$, as an illustrative application demonstrating the sensitivity to the light quark mass ratio parameter $Q$. Using the BESIII data for these two decays published in Ref.~\cite{BESIII:2017djm}, we obtain $Q=22.5(1.0)$, which is compatible with previous determinations and has a comparable uncertainty.

    The method can be further refined by including the $YX^2$ and $X^4$ terms, as done in Ref.~\cite{Escribano:2010wt}, which however are not available in the BESIII analysis in Ref.~\cite{BESIII:2017djm}.
    Furthermore, the treatment of final state interactions in the decays can be improved by using a dispersion framework~\cite{Isken:2017dkw,Akdag:2021efj} to reduce the theoretical uncertainty from unitarizing the NLO large $N_c$ ChPT amplitudes.

    BESIII has recently published a more thorough analysis of the $\eta'\to\eta\pi^0\pi^0$ decays \cite{BESIII:2022tas} with eight times more data. Although they include the $\pi\pi$ rescattering effect, they still lack the $YX^2$ and $X^4$ terms in their Dalitz plot distribution expansion. Nevertheless, once the full data set for the $\eta'\to\eta\pi^+\pi^-$ is available, the isospin-breaking parameter $Q$ can be extracted with a significantly reduced statistical uncertainty. 

    The method can also be applied to other reactions, such as decays into $J/\psi \pi\pi$ from higher charmonium(-like) states, as well as analogous decays in the bottomonium sector, to extract the IB effects therein. For such reactions with the initial state mass higher than open-charm (open-bottom) thresholds, the IB effects are expected to be more complicated since the isospin mass splittings of intermediate open-flavor mesons could play a crucial role. Using the unit disk distribution method to extract the IB effects in these reactions would be of great interest for the study of the isospin breaking dynamics in the heavy quark sector.

\begin{acknowledgements}
    We would like to thank Martin Hoferichter and Bastian Kubis for useful comments.
    This work is supported in part by the National Natural Science Foundation of China (NSFC) under Grants No.~12125507, No.~12361141819, No.~12447101, and No.~12405100; by the Chinese Academy of Sciences under Grant No.~YSBR-101; and by the National Key R\&D Program of China under Grant No. 2023YFA1606703.

\end{acknowledgements}

\appendix

\section{Isospin conserving amplitude}
\label{app:ICamp}

The ChPT Lagrangian density gives the dynamics of the pseudo-Nambu-Goldstone bosons in a nonlinear realization of the symmetry through the field $U(\phi)=\exp\left[i\frac{\sqrt{2}}{F}\phi\right]$, where 
\begin{equation}
   \phi 
   =\left(\begin{array}{ccc}\frac{\pi^0+C_q\eta+C_q'\eta'}{\sqrt{2}}&\pi^+&K^+\\\pi^-&\frac{-\pi^0+C_q\eta+C_q'\eta'}{\sqrt{2}}&K^0\\K^-&{\bar K^0}&-C_s\eta+C_s'\eta'\end{array}\right),
\end{equation}
and $F$ is the pion decay constant in the chiral limit. 
Here we have followed the two-mixing angle scheme for the neutral mesons~\cite{Kaiser:2000gs,Feldmann:1998vh,Feldmann:1998sh},
where the mixing constants are parameterized in the most general form as 
\begin{subequations}
   \begin{align}
       &C_q\equiv \frac{F}{\sqrt{3}\cos\left(\theta_8-\theta_0\right)}\left(\frac{\cos\theta_0}{f_8}-\frac{\sqrt{2}\sin\theta_8}{f_0}\right),\\
       &C_q'\equiv \frac{F}{\sqrt{3}\cos\left(\theta_8-\theta_0\right)}\left(\frac{\sqrt{2}\cos\theta_8}{f_0}+\frac{\sin\theta_0}{f_8}\right),\\
       &C_s\equiv \frac{F}{\sqrt{3}\cos\left(\theta_8-\theta_0\right)}\left(\frac{\sqrt{2}\cos\theta_0}{f_8}+\frac{\sin\theta_8}{f_0}\right),\\
       &C_s'\equiv \frac{F}{\sqrt{3}\cos\left(\theta_8-\theta_0\right)}\left(\frac{\cos\theta_8}{f_0}-\frac{\sqrt{2}\sin\theta_0}{f_8}\right).
   \end{align}
\end{subequations}
We will use the values of the couplings $f_{8/0}$ and the two mixing angles $\theta_{8/0}$ from Ref.~\cite{Guevara:2018rhj}.
The relevant operators in the NLO ($\order{\delta}$) Lagrangian of large $N_c$ ChPT are~\cite{Leutwyler:1996sa,Kaiser:2000gs,Escribano:2010wt}
\begin{eqnarray}\label{eq:L_4}
   \mathcal{L}_{(1)}&=&L_2\langle\partial_\mu U^\dagger\partial_\nu U\partial^\mu U^\dagger\partial^\nu U\rangle\nonumber\\&&+(2L_2+L_3)\langle\partial_\mu U^\dagger\partial^\mu U\partial_\nu U^\dagger\partial^\nu U\rangle\nonumber\\
   &&+L_5\langle\partial_\mu U^\dagger\partial^\mu U\left(U^\dagger\chi+\chi^\dagger U\right)\rangle\nonumber\\
   &&+L_8\langle U^\dagger\chi U^\dagger\chi+\chi^\dagger U\chi^\dagger U\rangle\nonumber\\
   &&-i\frac{F\Lambda_2}{6\sqrt{2}}\left(C_0\eta+C_0'\eta'\right)\langle U^\dagger\chi-\chi^\dagger U\rangle.
\end{eqnarray}

With these operators the IC amplitude for the $\eta'\to\eta\pi\pi$ decay is 
\begin{eqnarray}\label{eq:leading_amplitude}
   \mathcal{M}_{\eta'\to\eta\pi\pi}^{\rm IC}=\frac{c_{qq}}{F^2}\left[\frac{M_\pi^2}{2}+\frac{24L_8}{F^2}M_\pi^4+\frac{2}{3}\Lambda_2M_{\pi}^2\right.\hspace*{8ex}&&\nonumber\\+\frac{2(3L_2+L_3)}{F^2}\left(s^2+t^2+u^2-M_{\eta'}^4-M_\eta^4-2M_{\pi}^4\right)&&\nonumber\\
   \left.-\frac{2L_5}{F^2}\Delta_{\eta'\eta\pi}M_{\pi}^2\right] + \frac{c_{sq}}{F^2}\frac{\sqrt{2}}{3}\Lambda_2M_{\pi}^2,\qquad&&
\end{eqnarray}
where $M_\pi$ is the pion mass in the isospin limit, $c_{qq}=-2C_qC_q'$, $c_{sq}=C_q'Cs-C_qC_s'$ and $F=92.2$~MeV is the physical pion decay constant. The Mandelstam variables are defined as, taking the $\eta'\to\eta\pi^+\pi^-$ as an example, $s\equiv\left(p_{\pi^+}+p_{\pi^-}\right)^2$, $t \equiv\left(p_{\eta^{\prime}}-p_{\pi^{+}}\right)^2$, and $u \equiv\left(p_{\eta^{\prime}}-p_{\pi^{-}}\right)^2$; while the off-shell $\pi^0\to\pi^+\pi^-\pi^0$ amplitude reads
\begin{eqnarray}\label{eq:pi_to_3pi}
    &&\mathcal{M}_{\pi^0\to\pi^+\pi^-\pi^0}^{\rm IC}=\frac{1}{3F^2}\left(3s-\Delta_{\eta'\eta\pi}+M_\pi^2+\frac{64}{F^2}M_\pi^4L_8\frac{}{}\right.\nonumber\\
    &&\qquad-\frac{4}{F^2}\left\{6L_2\left[t(\Delta_{\eta'\eta\pi}-s-t)-M_{\eta'}^2M_\eta^2-M_\pi^4\right]\frac{}{}\right.\nonumber\\
    &&\qquad-3(2L_2+L_3)(s-M_{\eta'}^2-M_\eta^2)(s-2M_\pi^2)\nonumber\\
    &&\quad \left.\left.\frac{}{}-M_\pi^2L_5(12s-5\Delta_{\eta'\eta\pi})\right\} \right),
\end{eqnarray}
where $\Delta_{\eta'\eta\pi}=M_{\eta'}^2+M_\eta^2+2M_\pi^2$.

We account for $\pi\pi$ final state interactions with the $N/D$ unitarization method~\cite{Chew:1960iv,Oller:1998zr}
as of Ref.~\cite{Escribano:2010wt},
and the $\eta\pi$ rescattering ($t$- and $u$-channels) in these decays is negligible~\cite{Schneider:2009rz,Kubis:2009vu}.
In doing so, the total amplitude is expressed in terms of partial waves. For this process the relevant partial waves are those with $J=0,2$, since the two-pion system must have $I=0$ and higher angular momentum contributions are suppressed by more powers of momenta. The unitarized amplitude 
reads~\cite{Escribano:2010wt}
\begin{align}
   \mathcal{M}(s,t,u)=&\,\sum_J32\pi(2J+1)P_J(\cos\theta_\pi)\nonumber\\
   &\times\frac{\left.\mathcal{M}_J\right|_{\text{tree}}(s)}{1-16\pi B_0(s)\left.\mathcal{T}_J^0(s)\right|_{\text{tree}}},
\end{align}
where the $\mathcal{M}_J|_\text{tree}(s)$ is the partial-wave amplitude of total angular momentum $J$ at tree level, $\mathcal{T}^0_J(s)\big|_\text{tree}$ is the $I=0$ $\pi\pi$ scattering amplitude with total angular momentum $J$ at tree level, and 
\begin{equation} 
    16\pi^2B_0(s)=C-\rho(s)\log\frac{\rho(s)+1}{\rho(s)-1},
\end{equation}
where $\rho=\sqrt{1-4M_\pi^2/s}$ and $C$ is a constant.
Details of these partial-wave amplitudes can be found in Ref.~\cite{Escribano:2010wt}.

\section{Construction of the unit disk mapping}
\label{app:mapping}

In this appendix, we describe the construction of the mapping from the Dalitz plot to the unit disk. We first parameterize the boundary of the unit disk by dividing it into two segments $\left(D=D^{+} \cup D^{-}\right)$:
\begin{subequations}
\begin{align}
   & D^{+}=\left.\left\{\frac{y}{\sqrt{1-x^2}}=~~1~ \right| x \in[-1,1]\right\},\label{eq:Bound_disk_up} \\
   & D^{-}=\left.\left\{\frac{y}{\sqrt{1-x^2}}=-1~ \right| x \in[-1,1]\right\},\label{eq:Bound_disk_down}
\end{align}
\end{subequations}
where $x$ and $y$ are the usual Cartesian coordinates. The segment in Eq.~\eqref{eq:Bound_disk_up} is the upper part of the disk ($y\geq0$), while that in Eq.~\eqref{eq:Bound_disk_down} is the lower part ($y\leq0$). 

\begin{figure}[tb]
   \centering
   \includegraphics[width=0.45\linewidth]{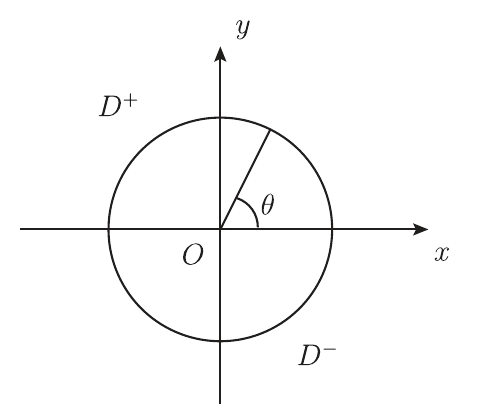}~~~
   \includegraphics[width=0.55\linewidth]{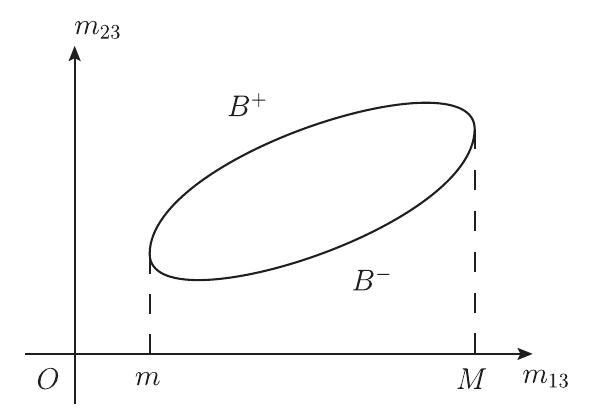}
   \caption{Left: the unit disk with its boundary divided into two segments, $D^+$ for $\sin\theta\geq0$ and $D^-$ for $\sin\theta\leq0$. Right: the two segments of the boundary of a conventional Dalitz plot where $B^+$ and $B^-$ correspond to $\cos\theta_{23}=1$ and $\cos\theta_{23}=-1$, respectively (see Eq.~\eqref{eq:B_function}). }
   \label{fig:standarddisc}
\end{figure}

The boundary can be parameterized by the angle subtended by each point of $D$ with respect to the $x>0$ axis, so that $\theta=0$ corresponds to the maximum value of $m_{12}$:
\begin{equation}\label{eq:x,y}
   x=\cos \theta, \quad
   y=\sin \theta, \quad \theta \in[0,2 \pi).
\end{equation}
Then the boundary $D$ depends only on $\theta$ (see the left panel of Fig.~\ref{fig:standarddisc}). 

The boundary of the conventional Dalitz plot is set to be a function of the invariant masses $m_{12}^2=(p_1+p_2)^2$ and $m_{23}^2=(p_2+p_3)^2$, constrained by kinematics. It is also divided into two segments $\left(B=B^{+} \cup B^{-}\right)$ (see the right panel of Fig.~\ref{fig:standarddisc}):
\begin{align}\label{eq:boundary_cond}
  B^{\pm}=\left.\left\{B\left(m_{12}^2, m_{23}^2\right)=\pm1~ \right| m_{12} \in[m_1+m_2, m-m_3]\right\},
\end{align}
where $B\left(m_{12}^2, m_{23}^2\right)$ is defined in Eq.~\eqref{eq:B_function} as the cosine of the angle between the three-momenta $\vec{q}_2^{\,*}$ and $\vec{q}_3^{\,*}$ in the c.m.\ frame of particles 1 and 2.
Furthermore, the equation of a Dalitz plot boundary, $[B\left(m_{12}^2, m_{23}^2\right)]^2=1$, can also be expressed as the Kibble cubic function:
$$\begin{aligned}
& s t u+2(m_1^2 m_2^2 m_3^2+p^2 m_1^2 m_2^2+p^2 m_2^2 m_3^2+p^2 m_3^2 m_1^2)= \\
&s(m_1^2 m_2^2\hspace{-.5mm}+\hspace{-.5mm}p^2 m_3^2)+t(m_2^2 m_3^2\hspace{-.5mm}+\hspace{-.5mm}p^2 m_1^2)+u(m_3^2 m_1^2\hspace{-.5mm}+\hspace{-.5mm}p^2 m_2^2),
\end{aligned}$$
where $s=m_{12}^2$, $t=m_{23}^2$, $u=m_{13}^2=m^2+m_1^2+m_2^2+m_3^2-m_{12}^2-m_{23}^2$ and $m$ is the mass of the initial particle.

It is easy to obtain the coordinates in the conventional Dalitz plot where $m_{12}$ has its maximum and minimum values, which we call, respectively, $\bm{a}=(a_1,a_2)$ and $\bm{b}=(b_1,b_2)$, through Eq.~\eqref{eq:B_function}; that is, $b_1=\left(m_{12}^2\right)_{\rm min}$, $a_1=\left(m_{12}^2\right)_{\rm max}$, and these conditions fix the second components of both vectors. 
We also define $\bm{c}^{\pm}=(\bm{a}\pm \bm{b})/2$, which is used to set the center of the disk and the orientation of the Cartesian coordinate system with respect to the Dalitz plot coordinate system.
In this way, $\bm{c}^-$ is a vector pointing from the center of the Dalitz plot to the point with the maximum value of $m_{12}^2$, and $\bm{c}^+$ is a vector pointing from the origin to the center.

Consider a mapping $S:[(m_1+m_2)^2,(m-m_3)^2]\times[0,2\pi) \to [(m_2+m_3)^2, (m-m_1)^2]$ with the following explicit form:
\begin{equation}
{\small
  S\big(m_{12}^2,\theta\big)=\left\{\begin{array}{ll} 
    \!\!\tan{\beta} \left(m_{12}^2-c_1^+\right)+c_2^+ & \text{for } \beta\neq\frac{\pi}{2},\frac{3\pi}{2}, \\[1mm] \!\!\text{Solution of } B\left(c^+_1,m_{23}^2\right)=~~1 & \text {for } \beta=\frac{\pi}{2}, \\[1mm] \!\!\text{Solution of } B\left(c^+_1,m_{23}^2\right)=-1 & \text {for } \beta=\frac{3\pi}{2},\end{array}\right.
}
\end{equation}
where $\beta = \theta+\alpha$, $\theta \in[0,2\pi)$, and $\alpha$ is the angle between the $m_{12}$ axis and $\bm{c}^-$. One has
\begin{align}
    \tan\alpha=\frac{c^-_2}{c^-_1} = \frac{m_1 m_3 - m m_2}{(m-m_3)(m_1+m_2)}.
\end{align} 
For the decays at hand, we choose the final state pion pair to be particles 1 and 2 and $\eta$ as particle 3. Thus, for both decays, we find $\tan\alpha=-1/2$, and $\alpha=-\arctan(1/2)$. 

Then one can construct a one-to-one mapping from the unit circle $D$ to the Dalitz plot boundary $B$, $\bm{f}: D \to B$, or equivalently,
\begin{equation}
\bm{f}: \theta \mapsto\left(m_{12}^2, m_{23}^2\right)\equiv\left(L(\theta), R(\theta)\right).
\end{equation}
Here, $L(\theta)$ is the solution to Eq.~\eqref{eq:Boundary_Soln}
with respect to $m_{12}^2$, and $R(\theta)=S\big(L(\theta),\theta\big)$. Thus, $\bm{c}^+$ gives the coordinates of the center of the disk in the invariant-mass coordinate system of the Dalitz plot and $\bm{c}^-$ the angle $\alpha$ between the $x$ and $m_{12}^2$ axes.

\begin{figure}
  \centering
  \includegraphics[width=0.9\linewidth]{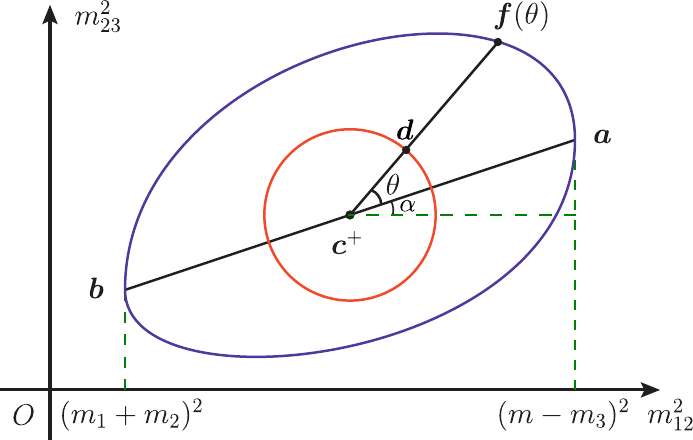}
  \caption{Mapping from a Dalitz plot to the unit disk, where $m$ is the mass of the initial state particle and $m_i~(i=1,2,3)$ are masses of the final state particles.}
  \label{fig:mapping_appd}
\end{figure}

Up to now, we have constructed a mapping from $D$ to $B$ with $\bm{f}(\theta)=\big(L(\theta), R(\theta)\big)$. Next we use the boundary mapping $\bm{f}$ to construct the mapping $\bm{F}:\left(m_{12}^2, m_{23}^2\right) \mapsto(r, \theta)$ from the conventional Dalitz plot to the unit disk. The procedure is as follows. If $\left|\bm{d}-\bm{c}^+\right|=0$, where $\bm{d}=\left(m_{12}^2, m_{23}^2\right)$ is any point belonging to the Dalitz plot (see Fig.~\ref{fig:mapping_appd}), then $r=0$; otherwise,
\begin{align}\label{eq:F_map}
  \cos \theta =\frac{\left(\bm{d}-\bm{c}^{+}\right) \cdot \bm{c}^{-}}{\left|\bm{d}-\bm{c}^{+}\right| \hspace*{.5ex}\left|\bm{c}^{-}\right|}, \quad
  r = \frac{\left|\bm{d}-\bm{c}^{+}\right|}{\left|\bm{f}(\theta)-\bm{c}^{+}\right|}.
\end{align}
For the division of the unit disk into bins, we use Cartesian coordinates: we first generate a mesh in the circumscribed square of the unit circle with lines parallel to the $x$ and $y$ axes in Eq.~\eqref{eq:x,y}. To do this, we generate 100 equidistant lines parallel to each axis, dividing the square into 10,000 equal-size bins. Afterwards we select those that fulfill $r\leq1$; after discarding the bins outside the unit circle we are left with 7,837 bins. The fraction of accepted bins is close to the ratio of the areas of the unit disk and the unit square, which is $\pi/4\approx78.54\%$. 

\section{Input Dalitz plot distribution parameters}
\label{app:parameters}

The Dalitz plot distribution parameters for the $\eta'\to\eta\pi^0\pi^0$ and $\eta'\to \eta\pi^+\pi^-$ decays extracted by the BESIII Collaboration are~\cite{BESIII:2017djm}
\begin{subequations}
\begin{align}
  &a_{\pi^0}=-0.087\pm0.009,\hspace*{1ex}b_{\pi^0}=-0.073\pm0.006,\nonumber\\&\hspace*{13ex}d_{\pi^0}=-0.074\pm0.004,\\
  &a_{\pi^\pm}=-0.056\pm0.004,\hspace*{0.5ex}b_{\pi^\pm}=-0.049\pm0.006,\nonumber\\&\hspace*{13ex}d_{\pi^\pm}=-0.063\pm0.004,
\end{align}
\end{subequations}
where their correlation matrices are reported to be 
\begin{equation}
   C_{\pi^0}=\left(\begin{array}{c|cc}&b_{\pi^0}&d_{\pi^0}\\\hline a_{\pi^0}&-0.495 &-0.273\\b_{\pi^0} &&0.273\end{array}\right),
\end{equation}
for the decay into neutral pions and 
\begin{equation}
   C_{\pi^\pm}=\left(\begin{array}{c|cc}&b_{\pi^\pm}&d_{\pi^\pm}\\\hline a_{\pi^\pm}&-0.417 &-0.239\\b_{\pi^\pm} &&0.292\end{array}\right),
\end{equation}
for the decay into charged pions. 

   \bibliography{refs.bib}

\end{document}